\documentclass[prl,twocolumn]{revtex4}
\usepackage{graphicx}
\usepackage{dcolumn}
\usepackage{bm}
\usepackage{textcomp}
\usepackage{float}
\usepackage[colorlinks]{hyperref}
\usepackage{epstopdf}

\begin{document}

\title{Stark effect and generalized Bloch-Siegert shift in a strongly driven two-level system}

\author{Jani Tuorila$^1$}
\author{Matti Silveri$^1$}
\author{Mika Sillanp\"a\"a$^2$}
\author{Erkki Thuneberg$^1$}
\author{Yuriy Makhlin$^{2,3}$}
\author{Pertti Hakonen$^2$}

\affiliation{$^1$Department of Physics, University of Oulu, FI-90014, Finland \\
$^2$Low Temperature Laboratory, Aalto University, P.O. Box 15100, FI-00076 AALTO, Finland\\
$^3$Landau Institute for Theoretical Physics, Kosygin st. 2, 119334, Moscow, Russia}

\date{\today}

\begin{abstract}
A superconducting qubit was driven in an ultrastrong fashion by an oscillatory microwave field, which was created by coupling via the nonlinear Josephson energy. The observed Stark shifts of the `atomic' levels are so pronounced that corrections even beyond the lowest-order Bloch-Siegert shift are needed to properly explain the measurements. The quasienergies of the dressed two-level system were probed by resonant absorption via a cavity, and the results are in agreement with a calculation based on the Floquet approach.
\end{abstract}


\maketitle

The characteristics of matter and light become intertwined upon interaction. The interconnection of the two can be observed in atomic and optical physics by setting atoms inside mirrored cavity resonators, whereby coupling the zero-point vibrations of the field to atomic transitions. Pioneering studies of quantum physics have been carried out in the scheme (see, e.~g.,~\cite{Cohen}).

Recently, an increasing emphasis has been put on the study of analogous physics in a setting of electromagnetic modes interacting with discrete systems, but in a solid-state environment. Such quantum few-level systems, or, artificial atoms, have been implemented based on quantum dots and superconducting Josephson qubits~\cite{nakamura,Vion02,Schoelkopf08}.

Apart from zero-point vibrations in a cavity, natural or artificial atoms can be coupled to a driven laser field. One of the effects of the field is the dynamic (ac) Stark shift~\cite{AT,Scully,Schuster,baur09,EIT} of the energy levels. For off-resonant driving, the shift scales linearly in the number of quanta in the field. An additional correction, the Bloch-Siegert shift, appears for an oscillating, rather than circularly polarized field~\cite{BS,Abragam}. A strong drive with the Rabi frequency $\Omega_R$ becoming a sizable fraction of the atomic level spacing is needed to reveal the realm of such delicate phenomena, which poses a challenge for experimenting with real atoms. A good understanding of the physics is important also in the sense that the artificial systems have been actively investigated due to their promise of setting up quantum information processing.

In the present work, we investigate the energy levels of an artificial two-level system driven by an oscillatory field originating from a harmonic drive via the Josephson energy. For the purpose, we have developed a qubit-resonator setup, where matrix elements allow to carry out the measurement over the entire excursion covered by the drive. We measured large Stark shifts of the qubit level spacing unseen in atomic systems. The shifts are found to exhibit unconventional and to some extent nonmonotonic dependence on the field amplitude. This work is the first observation of Bloch-Siegert type of correction in driven systems other than atomic systems~\cite{BSrydberg} or magnetic resonance~\cite{Abragam}. Unlike all but few previous experiments~\cite{BShigh}, we have to go beyond the usual lowest-order such correction in order to account for the data. This work differs from the previous studies on strong drive with superconducting or quantum dot qubits, which focused on the Landau-Zener-St\"uckelberg (LZS) effect~\cite{LZSM,rydbergLZ}, and where the coupling of the drive to the qubit Hamiltonian was linear~\cite{Nakamura01,feigelman,Oliver05,Sillanpaa05,kayanuma06,Wilson07,Ashhab07,Izmalkov08,baur09,EIT,sun09,Shevchenko09,qdotLZ}.


\begin{figure}[tH]
\centering
\includegraphics[width=0.8\linewidth]{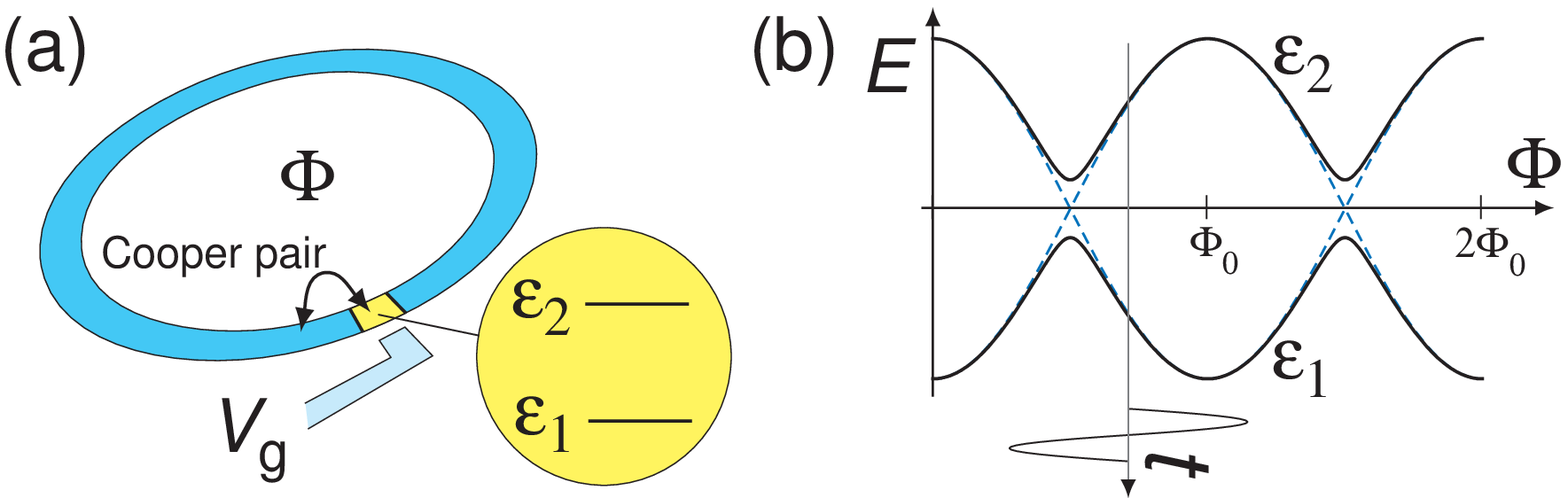}\\
\includegraphics[width=0.8\linewidth]{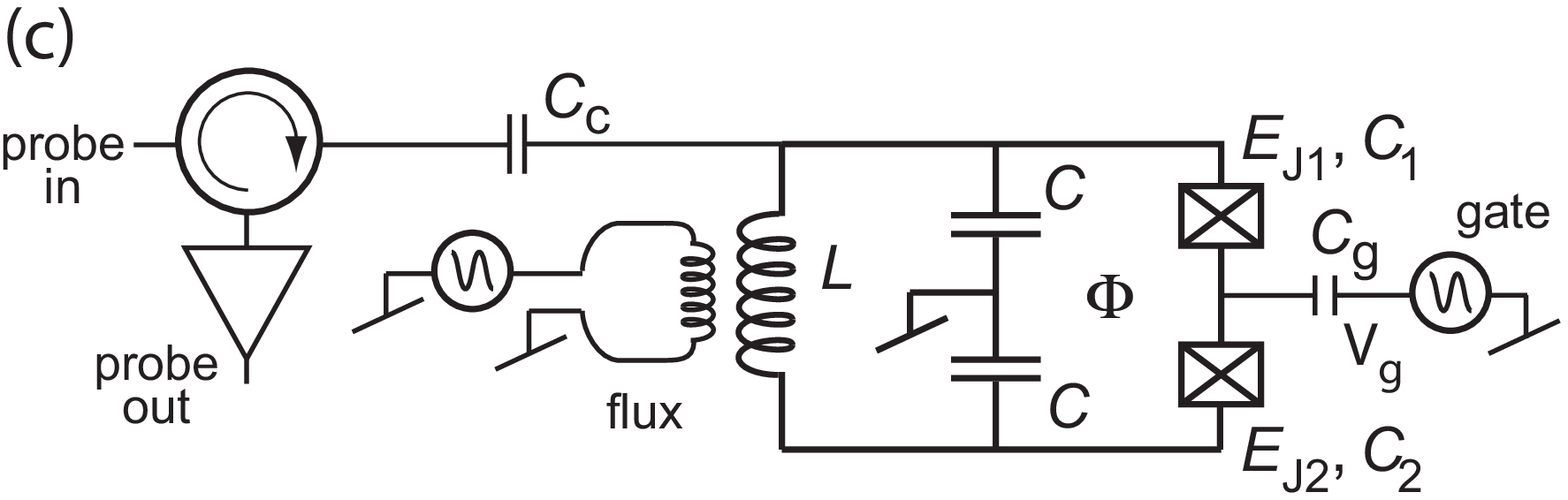}
\caption{Schematic of the experimental setup. (a) The single-Cooper-pair transistor qubit consists of a superconducting loop interrupted by two Josephson junctions that separate a small island. The island has two relevant charge states corresponding to $n$ or $n+1$ Cooper pairs; (b) Except near the anticrossings, the energy eigenvalues depend nearly sinusoidally on the applied flux $\Phi$. The flux is driven likewise sinusoidally in time; (c) A more detailed view of the circuit shows the microwave reflectometry readout via an $LC$-resonator formed by on-chip lumped circuit elements \cite{Gunnarsson08}.}\label{schema}
\end{figure}

We use as the two-level system a single-Cooper-pair transistor (SCPT)~\cite{nakamura,Vion02}, a superconducting qubit which consists of two small-area tunnel junctions with the Josephson energies $E_{J1}$ and $E_{J2}$, see Fig.~\ref{schema}. The phase difference across the SCPT can be tuned by means of a magnetic flux $\Phi$ applied through a superconducting loop. In order to minimize background charge noise, the gate voltage is adjusted so that two charge states differing by one Cooper pair are degenerate~\cite{Vion02}. Neglecting higher charge states, the Hamiltonian of the SCPT written in the charge basis consists of the Josephson coupling energies,
\begin{equation}\label{Hq}
H_q=-\frac{E_{J0}}{2}\Big[\cos\Big(\frac{\pi\Phi}{\Phi_0}\Big)\sigma_x-d\sin\Big(\frac{\pi\Phi}{\Phi_0}\Big)\sigma_y\Big] \,.
\end{equation}
The parameters are the total Josephson energy $E_{J0} = E_{J1}+E_{J2}$, the asymmetry $d=(E_{J1}-E_{J2})/E_{J0}$, and the flux quantum $\Phi_0=h/2e$. The energy eigenvalues $E_J(\Phi)=E_{J0}\sqrt{\cos^2(\pi\Phi/\Phi_0)+d^2\sin^2(\pi\Phi/\Phi_0)}$ of $H_q$~(\ref{Hq}) are depicted in Fig.~\ref{schema}b.

The applied flux $\Phi=\Phi_b+\Phi_L\cos(\omega_Lt)$ consists of a static bias $\Phi_b$ and of a time-dependent part with the amplitude $\Phi_L$ which is analogous to the intense laser field in atomic physics. The on-chip flux coil was designed to have a large mutual inductance of $\sim 5$~pH to the superconducting loop in order to achieve a desired fast control of the flux bias over the span of several $\Phi_0$. Owing to the nested sinusoidal time dependence of the energy, as well as unequal coupling to the different $\sigma_{x,y,z}$, the driving field substantially deviates from a circular polarization, and novel phenomena appear.

\begin{figure}[tH]
\centering
\includegraphics[width=0.8\linewidth]{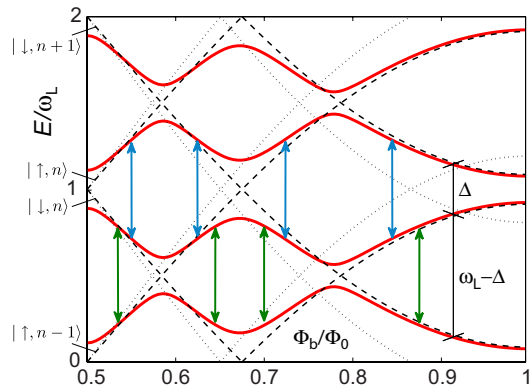}
\caption{Calculated quasienergies as a function of the bias flux $\Phi_b$ with the experimental parameters. The fully dressed states (red lines) are the eigenvalues of the Floquet matrix. The quasienergies repeat periodically in energy with period $\hbar\omega_L$. The level splittings are denoted by
$\hbar\Delta$ and $\hbar(\omega_L-\Delta)$.
The vertical green and blue arrows denote transitions induced by  3.5~GHz probe field.
We also show the quasienergies of the bare states ($\Phi_L=0$, dotted lines) and longitudinally dressed states ($\Omega_k=0$, dashed lines).}
\label{dressedstates}\end{figure}

In the following we find the coupled states of the qubit and the field. This leads to the picture of the dressed states formed by the qubit and the light field. The spectrum is obtained as quasienergies, which can be considered as the characteristic energies of the combined system of the qubit and the field~\cite{Cohen}. They repeat periodically at intervals $\omega_L$ as depicted in Fig.~\ref{dressedstates}.

Assuming $d$ small, we diagonalize the first term in $H_q$ (\ref{Hq}), replacing $\sigma_x\to\sigma_z$, $\sigma_y\to\sigma_x$. The eigenstates of this term are called ``bare'' below. For simplicity, we give explicit formulas in the case where we keep only the 0th and 1st harmonics of the driven Hamiltonian~(\ref{Hq}). The numerical calculations, however, are done for the full Hamiltonian. To account for the temporal variation of the longitudinal field, we transform to a rotating frame. The transformation is obtained by $U = \exp\big(-i\sigma_zA\sin(\omega_L t)/2 \hbar\omega_L\big)$, where $A=2E_{J0}J_1(\pi\Phi_L/\Phi_0)\sin(\pi\Phi_b/\Phi_0)$. We define $\hbar \omega_0 =- E_{J0}J_0(\pi\Phi_L/\Phi_0)\cos(\pi\Phi_b/\Phi_0)$, $B=E_{J0}dJ_0(\pi\Phi_L/\Phi_0)\sin(\pi\Phi_b/\Phi_0)$ and $\Omega_k = [B+k\omega_Ld\cot\big(\pi\Phi_b/\Phi_0\big)] J_k(A/\hbar\omega_L)$ and find
\begin{equation}\label{H0}
H_{\rm ld} = \frac{\hbar}{2}\Big[ \omega_0 \sigma_z + \sum_{k=-\infty}^{\infty}\Big(\Omega_k\sigma_+ +\Omega_{-k}\sigma_-\Big)e^{i k\omega_L t}\Big].
\end{equation}

The strongest effect of the ``laser'' field is a modification of the qubit splitting to $\hbar\omega_0$ by a factor of $J_0(\pi\Phi_L/\Phi_0)$. This can be understood as a rectification of the drive by the qubit nonlinearity which effectively shifts the bias point, as seen by the transition from the dotted to dashed lines in Fig.~\ref{dressedstates}. The positions of the resonances (level crossings), hence become shifted, cf.~Fig.~\ref{dressedstates}.

Apart from that, the $\Omega_k$ terms describe transverse couplings induced by the time-dependent laser field. To take that into account, we consider $H_{\rm ld}$~(\ref{H0}) as a matrix operating on the longitudinally dressed states $\left|\sigma,n\right\rangle$ ~\cite{Nakamura01,Wilson07}. They are labeled by a qubit (`spin') index $\sigma=\uparrow,\downarrow$ and a Floquet (`photon') index $n$. The couplings have an effect of opening a gap at the $k$th anticrossing. This shifts the quasienergies from their uncoupled values (the ac Stark shift~\cite{AT}). For very weak couplings, $\Omega_k\ll\omega_L$, it is enough to keep only the couplings between resonant levels in the Floquet picture (rotating-wave approximation, RWA). RWA allows for analytic solutions and gives the Stark shift of $\sqrt{\delta_L^2+\Omega_k^2}-|\delta_L|$, where $\delta_L$ is the detuning from resonance. Among other approaches, RWA was sufficient for description of recent LZS experiments on superconducting qubits \cite{Nakamura01,Oliver05,Wilson07}.

\begin{figure}[tH]
\centering
\includegraphics[width=0.98\linewidth]{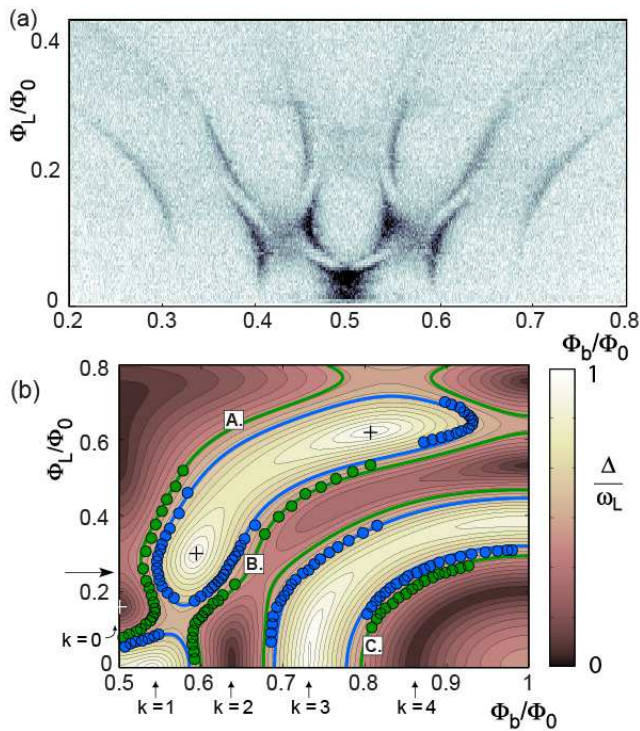}
\caption{(a) The measured resonance absorption of the probe signal plotted in the $\Phi_b-\Phi_L$ plane. Dark corresponds to higher absorption. The vertical scale is obtained by comparison to theory; (b) The landscape of the quasienergy splitting $\Delta$ calculated from the Floquet matrix of Hamiltonian (\ref{H0}) with the experimental parameters. The $k$-photon Rabi resonances appear as light (odd $k$) or dark (even $k$) tracks that are marked by index $k$ at small $\Phi_L$ and curve to the right with increasing $\Phi_L$.
Fig.~\ref{dressedstates} shows a cut along the horisontal line denoted by the black arrow. The blue and green lines denote the resonances in Eq.\ (\ref{rescond}). The solid dots are the experimental resonances picked up from panel (a), or from the measured phase shift (two rightmost lines).}\label{fig3}
\end{figure}

In our case, $\Omega_k/\omega_L$ is closer to 1, depending on the dc and ac bias, and thus a few orders of the perturbative expansion in this parameter beyond RWA should be accounted for (the second order, similar to the Bloch-Siegert contribution, and beyond).
Adding to $H_{\rm ld}$ (\ref{H0}) the photon energies, we solve numerically for the eigenvalues and eigenstates of the relevant part of the infinite Floquet matrix~\cite{Shirley65}. The calculated quasienergy splitting $\Delta$ for the drive amplitude $\Phi_L = 0.26 \, \Phi_0$ is plotted in Fig.~\ref{dressedstates}. Over the full ($\Phi_b,\Phi_L$) parameter space, the splitting is shown by contours in Fig.~\ref{fig3}(b).

In Fig.~\ref{fig3}(b) the Rabi resonances appear as light and dark tracks, corresponding to $\Delta/\omega_L\approx 1$ or $0$, respectively. These resonances start as vertical lines at small $\Phi_L$, but curve to the right with increasing $\Phi_L$. The curving is a consequence of approaching the first zero of $J_0(\pi\Phi_L/\Phi_0)$ at $\Phi_L\approx 0.77\Phi_0$, which implies a dynamic collapse of the qubit. This effect of the nonlinear longitudinal driving bears similarity to the coherent destruction of tunneling \cite{CDT} by transverse driving. It is visible as the dark horizontal band around $\Phi_L\approx 0.75\Phi_0$ in Fig.~\ref{fig3}(b). The plot also shows several conical points (crosses), where the energy values are degenerate ($\Delta/\omega_L=1$ or $0$ exactly)~\cite{LL}.

The experiments were performed in a dilution cryostat at a temperature of 30 mK. We first measured the sample parameters independently on the strong-drive experiments. Especially near the charge-flux degeneracy point, detailed mapping of the energy landscape is required in order to later obtain accurate comparison to the expected frequency shifts. The values of the total Josephson energy $E_{J0}/h = 27.0$~GHz and $d = 0.19$ were obtained by microwave spectroscopy at a very small drive, which directly yields the bare qubit's level spacing by making a fit to numerically evaluated energies taking into account 10 charge states. The ratio $E_{J0}/E_C = 8.0$ was obtained as a fit to the ground-state inductance over the $V_g$-$\Phi_b$ plane. Here, $E_C = e^2/2(C_1 + C_2)$, and the parallel junction capacitances are $C_1$ and $C_2$. The ground-state response agreed with the spectroscopic measurement of $E_{J0}$ and $d$.

Instead of measuring directly the excited state population \cite{Oliver05}, we probe the quasienergies with a weak $\mu$w-signal. The probe is produced by coupling the qubit with an $LC$-resonator via the total flux $\Phi$ in the loop (Fig.\ \ref{schema}c), and the information is encoded in the exchange of energy with the qubit, or in the dispersive frequency shift. The resonator is formed by the inductance of the superconducting loop, $L \sim 410$ pH, and by the lumped element capacitors $C \sim 10$ pF made out of Al oxide between the Al films. Since the $LC$ frequency $\sim 3.5$~GHz is smaller than the minimum level spacing $d E_{J0} \sim 5.1$~GHz of the bare qubit, the qubit-resonator interaction does not noticeably influence the energies.

For measurement, the circuit is excited at a frequency close to the resonance frequency, and the phase and amplitude of the reflected signal are recorded. We developed the following semiclassical description of the measurement. The resonator coupling is equivalent to adding a resonator-induced flux $\Phi_P(t)=\Phi_P\cos(\omega_P t)$ into $\Phi= \Phi_b + \Phi_L(t) + \Phi_P(t)$. We assume that $\Phi_P\ll \Phi_0$ so that it does not perturb the dressed states significantly. Transforming to the basis of longitudinally dressed states,  the Hamiltonian is $H=H_{\rm ld}+H_P$, where  $H_{\rm ld}$ is given in (\ref{H0}),
\begin{equation}\label{probe}
H_P = \frac{\pi\Phi_P}{2\Phi_0}\cos(\omega_P t)\sum_{n=-\infty}^{\infty}\Big(\delta_n\sigma_+ +\delta_{-n}\sigma_-\Big)e^{i n\omega_L t},
\end{equation}
We have defined $\delta_n=\hbar d\,(\omega_0+n\omega_L)J_n(A/\hbar\omega_L)$.
Assuming small $\Phi_P$, the probe Hamiltonian (\ref{probe}) can be treated as a small perturbation.
Besides the perturbative analysis, we have run a full simulation on the reflection measurement. For this purpose, we have derived the equations of motion for our circuit using quantum network theory \cite{Yurke84Devoret95,Tuorila09}. The resonator is treated classically, and it is coupled to Bloch equations describing the qubit.

\begin{figure}[tH]
\centering
\includegraphics[width=0.98\linewidth]{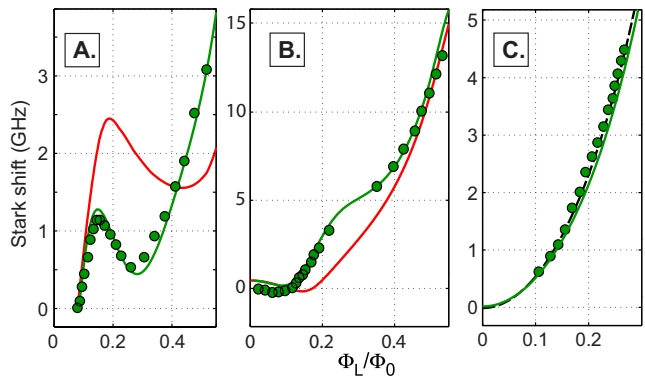}
\caption{The shifts of the spectral lines due to the qubit driven by a strong field. The data points are from Fig.~\ref{fig3}. The theoretical curves were produced with the rotating wave approximation in the adiabatic basis (red lines), or with the full numerical result (green). In (c), the dashed line is the analytical solution for the resonance condition, $(k \omega_L + \omega_P ) \{ \left[ J_0(\pi\Phi_L/\Phi_0) \right]^{-1} - 1 \}$, $k = 3$.}\label{fig4}
\end{figure}

In the dispersive regime, where the resonator is substantially detuned from the qubit, the resonator frequency changes according to the Josephson inductance of the qubit, which was used to obtain the parameters of the qubit in the ground state. With strong drive, however, the dispersive signal becomes overwhelmed by the absorptive response. Because the  probe frequency $\omega_P$ is smaller than the laser frequency $\omega_L$ ($\omega_L/2\pi = 6.11$~GHz and $\omega_P/2\pi = 3.5$~GHz), we can see two different transitions in the absorptive measurement. These are the closest states, corresponding to energy differences $\hbar\Delta$ and $\hbar(\omega_L-\Delta)$ in Fig.~\ref{dressedstates}a.  The resonance conditions are
\begin{equation}\label{rescond}
\Delta =\quad
\omega_P\ \textrm{(blue)} \quad\textrm{or}\quad \omega_L-\omega_P\ \textrm{(green)}
\,,
\end{equation}
where the color coding is related to Figs.~\ref{dressedstates} and \ref{fig3}a. They can be interpreted as the lowest transition in a fluorescent triplet, and the transition between Rabi states \cite{Cohen}.

In the $\Phi_b$-$\Phi_L$ plane, the resonance conditions appear as contours (Fig.~\ref{fig3}b). The matrix elements for the transitions are determined by $H_P$ (\ref{probe}). Since they are nonzero almost everywhere, accurate mapping of the energy landscape is possible. The resonant energy flow~\cite{Gunnarsson08} from the resonator to the qubit can be seen as increased absorption as well as a phase shift in the reflection measurement. The former is shown in Fig.~\ref{fig3}a. The measured locations of the absorption maxima follow closely the resonance lines (\ref{rescond}) as seen by overlaying them with the theory in Fig.~\ref{fig3}b. The full simulation of the measurement is in good agreement with the measured reflection.

One can also extract from Fig.~\ref{fig3} the actual Stark shift of the spectral line. The shift from the undriven case is illustrated in Fig.~\ref{fig4} for the three resonances labeled A-C in Fig.~\ref{fig3}. The overall trend is that the shift grows with the drive amplitude as $\left[ J_0(\pi\Phi_L/\Phi_0) \right]^{-1} - 1$ due to the rectification, which is obtained by inverting the effective qubit splitting $\omega_0$ for comparing to the original splitting. Moreover, the `Rabi' gap $\Omega_k$ modulates the shift near the $k$th resonance. Its Bessel-type dependence on the drive gives rise to the nonmonotonic Stark shift in Fig.~\ref{fig4}.
For the curve C the relevant coupling $\Omega_3$ is weak near $\Phi_b/\Phi_0=1$ and can be neglected. It is hence enough to account for the rectified level splitting $\omega_0$ in this region, cf.~Fig.~\ref{fig4}c.

The data display a good agreement to the full theory calculations (green). We also show a comparison to the RWA calculation (red), which clearly fails to explain the data. For the resonance of Fig.~\ref{fig4}a, higher order corrections even up to fourth order still substantially deviate from the data. The difference between the full calculation and the RWA can be considered a generalized Bloch-Siegert shift.

To conclude, we have measured and theoretically analyzed a superconducting qubit strongly driven via the Josephson energy. The spectral lines of the qubit experience a remarkably strong Stark shift up to the original level spacing. In order to obtain a good agreement between experiment and theory, we have to consider higher orders beyond the rotating-wave approximation which amounts to including a generalized Bloch-Siegert shift.

\begin{acknowledgments}
This work was financially supported by the Magnus Ehrnrooth foundation, the Academy of Finland, by the European Research Council (grant No.~FP7-240387), and the Dynasty foundation. We thank Sahel Ashhab for useful discussions.
\end{acknowledgments}

\end{document}